\documentclass[%
 reprint,
 amsmath,amssymb,
 aps,
]{revtex4-2}

\usepackage{graphicx}
\usepackage{dcolumn}
\usepackage{bm}
\usepackage{xcolor}

\begin{document}

\preprint{APS/123-QED}

\title{Control over cavity exciton polaritons in monolayer semiconductors}

\author{Zhi Wang}
\thanks{These authors contributed equally}%
\author{Li He$^*$}%
\email{heli1@sas.upenn.edu}
\author{Bumho Kim}%
\author{Bo Zhen}%
 \email{bozhen@sas.upenn.edu} 
\affiliation{%
 Department of Physics and Astronomy, University of Pennsylvania, Philadelphia, PA 19104, USA 
}%




\date{\today}

\begin{abstract}
Integrating two-dimensional van der Waals materials with optical cavities has revealed a fascinating platform to study exciton-polariton physics. 
Manipulating exciton polaritons often requires external control over the electrical and optical properties of materials. 
Here we demonstrate the electrical control of 2D exciton polaritons by strongly coupling a transition metal dichalcogenides (TMD) heterostructure to a photonic crystal nanocavity. 
Through precise control of the doping level in the TMD monolayers using electrostatic gating, we demonstrate a wide range of tunability in the exciton oscillator strength and hence the exciton photon hybridization. 
This tunability leads to the demonstration of a continuous transition from weak to strong coupling regime, as manifested by the disappearance and emergence of exciton polaritons, showcasing the versatility of our approach. 
Our work paves the way to further exploring nonlinear and quantum exciton polaritons in 2D materials and their device applications.

\end{abstract}

\maketitle



The coherent coupling between light and matter within an optical cavity results in the formation of quasiparticles known as polaritons, which exhibit dual characteristics of both light and matter. 
These polaritons serve as the foundation for a plethora of intriguing classical and quantum phenomena across diverse material systems and frequencies, including polariton lasing\cite{deng2003polariton}, polariton blockade\cite{delteil2019towards,munoz2019emergence} and polariton Bose-Einstein condensation\cite{byrnes2014exciton}. 
Recent advancements in two-dimensional transition metal dichalcogenides (TMDs) and their integration with photonic structures open new avenues for investigating exciton polaritons at the atomic scale. 
One distinctive feature of 2D TMD excitons is their valley degree of freedom and selective coupling to photon helicity, governed by the valley-dependent optical selection rules\cite{xiao2012coupled}. 
This internal valley degree of freedom, when applied to hybrid light-matter systems, have led to the realization of valley polarized exciton polaritons\cite{chen2017valley}. 
Furthermore, excitons in $K$ and $K'$ valleys are related by time-reversal operation. 
As a result, lifting the valley degeneracy in 2D exciton polariton systems could further enable the exploration of topological exciton polaritons with broken time-reversal symmetry at optical frequencies\cite{he2023polaritonic}. 
Another intriguing aspect of exciton physics in 2D TMDs arises when two monolayers are twisted and stacked to form bilayer heterostructures\cite{seyler2019signatures,jin2019observation}. 
Notably, the resulting moiré superlattices give rise to intralayer moiré excitons, characterized by significantly enhanced nonlinearity compared to their monolayer counterparts due to exciton blockade. 
As a result, the strong hybridization between moiré excitons and cavity photons can yield polaritons with substantial nonlinearity\cite{zhang2021van}, providing an interesting platform for the investigation of strongly interacting exciton polaritons.

2D exciton polaritons have been extensively explored in various photonic systems, ranging from \textcolor{black}{vertically stacked} dielectric Fabry-Perot cavities\cite{liu2015strong} and photonic crystal slabs\cite{zhang2018photonic,chen2021metasurface} to plasmonic lattices\cite{liu2016strong}.
\textcolor{black}{Notably, most optical resonators used in prior studies are relatively large and bulky, which leads to spatially extended polariton modes that are spread out over a large area.}
However, for many important applications where strong polariton nonlinearity is required, it is crucial to confine polaritons to the subwavelength scales \textcolor{black}{to achieve strong polariton-polariton interaction\cite{rosser2022dispersive}.} 
This is exemplified by several recent experimental work that employed semiconductor microcavities to confine polariton modes within a compact area ($\sim \lambda^2$),
\textcolor{black}{ enhancing polariton nonlinearity approaching the single particle level\cite{delteil2019towards,munoz2019emergence}.} 

Another challenge that often hinders the performance of 2D exciton-polariton systems is the sensitivity of exciton properties to the surrounding environment, such as charged impurities in various optical substrates\cite{burson2013direct}, and to the charge doping level in the exfoliated monolayers.
\textcolor{black}{For example, it has been shown that the spectral features of excitons in 2D monolayers can be precisely manipulated by adjusting the doping level through back-gated devices\cite{ross2013electrical,li2021refractive}.}
\textcolor{black}{However, these gate-tunable devices frequently involve metallic materials, like graphene, for the bottom gates. Such metals can cause significant optical absorption, which may spoil the optical resonances when integrated with optical cavities\cite{gu2019room}. 
Therefore, incorporating electrostatic tuning mechanisms into 2D TMD-cavity systems without compromising their optical performance remains a significant experimental hurdle.}

\begin{figure}[t]
    \centering
    \includegraphics[width=1\columnwidth]{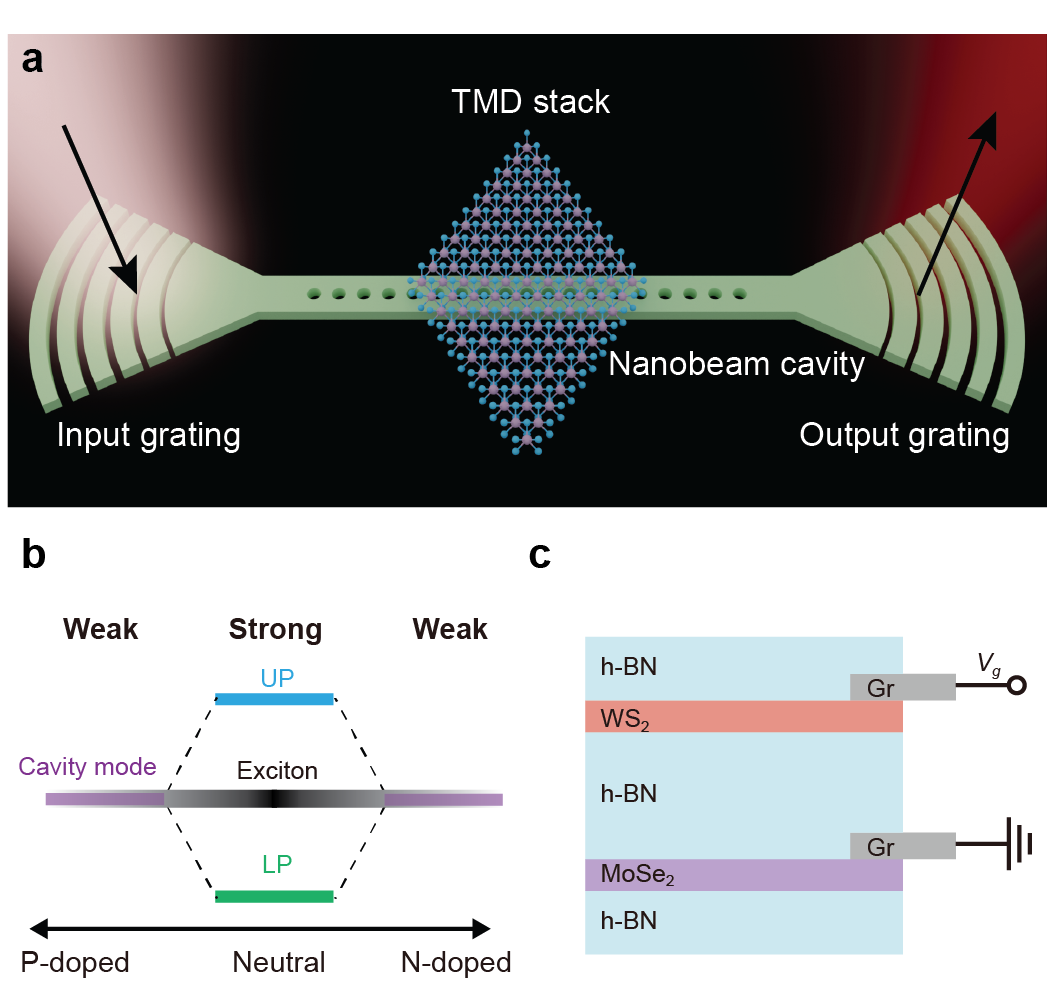} 
    \caption{\label{fig:epsart} Gate-tunable exciton-photon hybridization. a, Schematic of the TMD-PhC nanocavity. b, Energy level diagrams of the system at different doping levels. The continuous decrease of the doping concentration in the TMD monolayer results in an enhanced exciton-photon hybridization, leading to the emergence of upper and lower polariton states. c, Schematic of the TMD stack. The MoSe$_2$ and WS$_2$ monolayers are separated by 10 nm thick hBN layer and are encapsulated by $\sim$ 5 nm thick hBN layers from both top and bottom.}
\end{figure}

Here we demonstrate the capability to electrically control 2D exciton polaritons in a TMD heterostructure integrated in a planar photonic circuit (Fig. 1a).
\textcolor{black}{By employing a gate-tunable device that utilizes a WS$_2$ monolayer as a transparent top gate, we achieve substantial control over the exciton oscillator strength, allowing for a significant modulation of their optical response.}
\textcolor{black}{This strong modulation of exciton resonances within an optical cavity facilitates a continuous transition between weak and strong exciton-photon coupling regimes, as evidenced by the disappearance and reappearance of exciton polaritons (Fig.~1b).}
\textcolor{black}{Remarkably, the photonic crystal nanobeam cavity we employed boasts an extremely compact mode area of $\sim (\lambda/n)^2$, which is several orders of magnitude smaller than those explored in the prior studies.}
The reduced polariton size is anticipated to substantially enhance the effective photon-photon interaction strength, mediated by the exciton components.
\textcolor{black}{Furthermore, unlike previous studies that relied on free-space coupling to interact with exciton-polariton systems, our planar photonic architecture permits the separate investigation of excitons and polaritons. }
\textcolor{black}{Using this unique capability, we independently measured the photoluminescence of excitons and polaritons, thereby demonstrating the strong coherent coupling between excitons and cavity photons in the hybrid system. 
This method provides a distinct advantage over other coupling techniques.}

\begin{figure*}[t]
    \centering
    \includegraphics[width=2\columnwidth]{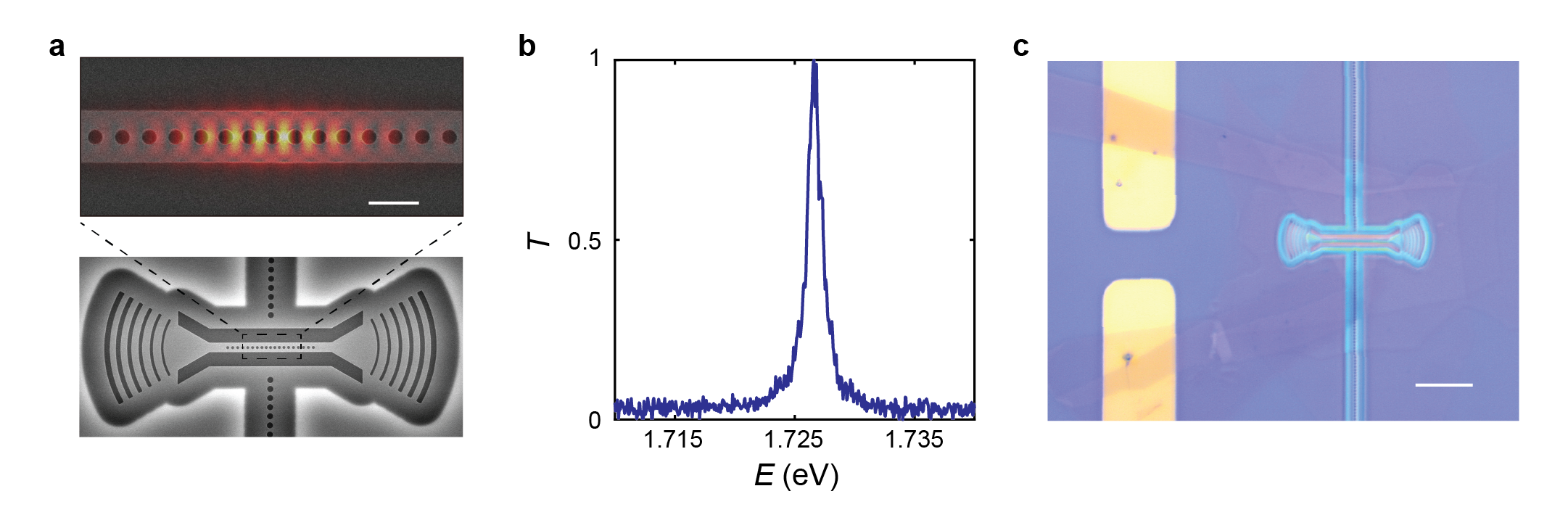}    \caption{\label{fig:wide} Coupled TMD-PhC nanocavity. a, Scanning electron microscope image of the photonic crystal nanobeam cavity and the simulated cavity mode profile. Scale bar, 500 nm. b, Optical transmission spectrum of the bare cavity measured through grating couplers. The bare cavity resonance has a loaded quality factor of 1200. c, Optical microscope image of the coupled TMD-PhC nanocavity. Scale bar, 10 $\mu$m.}
\end{figure*}

\textcolor{black}{Our TMD stack is schematically shown in Fig.~1c, which consists of MoSe$_2$ and WS$_2$ monolayers separated by ~$\sim$10 nm thick of hBN.
The WS$_2$ monolayer serves as a transparent top gate, avoiding significant optical absorption near the energy of MoSe$_2$ excitons.}
To demonstrate strong exciton-photon coupling, we fabricated a silicon nitride (Si$_3$N$_4$) photonic crystal nanobeam cavity, which is suspended above the SiO$_2$ substrate (see SI for more information). 
Light is coupled into and out of the nanobeam cavity using a pair of grating couplers at the two ends.
To facilitate gas release during the dry transfer process, an array of circular holes is patterned adjacent to the nanobeam cavity. 
\textcolor{black}{Scanning electron microscope images of the nanobeam cavity are shown in Fig.~2a, before transferring the TMD stack on top.}
\textcolor{black}{The transmission spectrum of the bare cavity is measured via the grating couplers (Fig.~2b), featuring a cavity resonance at 1.727 eV.}  
The quality factor of the bare cavity ($Q_\text{bare} \sim 1200$) is mostly determined by its radiative coupling to the waveguides on both sides, which can be readily adjusted by varying the number of holes employed. 
The TMD stack is then transferred onto the photonic chip with pre-patterned gold electrodes using the standard dry transfer technique (Fig.~2c).
The presence of the TMD stack induces a red-shift in the cavity resonance due to the change in the dielectric environment, aligning it with the exciton resonance of MoSe$_2$.
\textcolor{black}{Meanwhile, introduction of the TMD stack also lowers the loaded cavity quality factor to $Q_{\rm loaded}\sim 600$, owing to the reduced band gap confinement (see SI for more information).}
\textcolor{black}{Our design yields a cavity linewidth that is comparable to the MoSe$_2$ exciton linewidth of a few meV\cite{cadiz2017excitonic}.}
Consequently, it results in a narrow polariton linewidth when excitons are strongly coupled to the cavity resonance, while simultaneously ensuring robust optical readout signal through the coupling waveguides.

\begin{figure*}
    \centering
     \includegraphics[width=2\columnwidth]{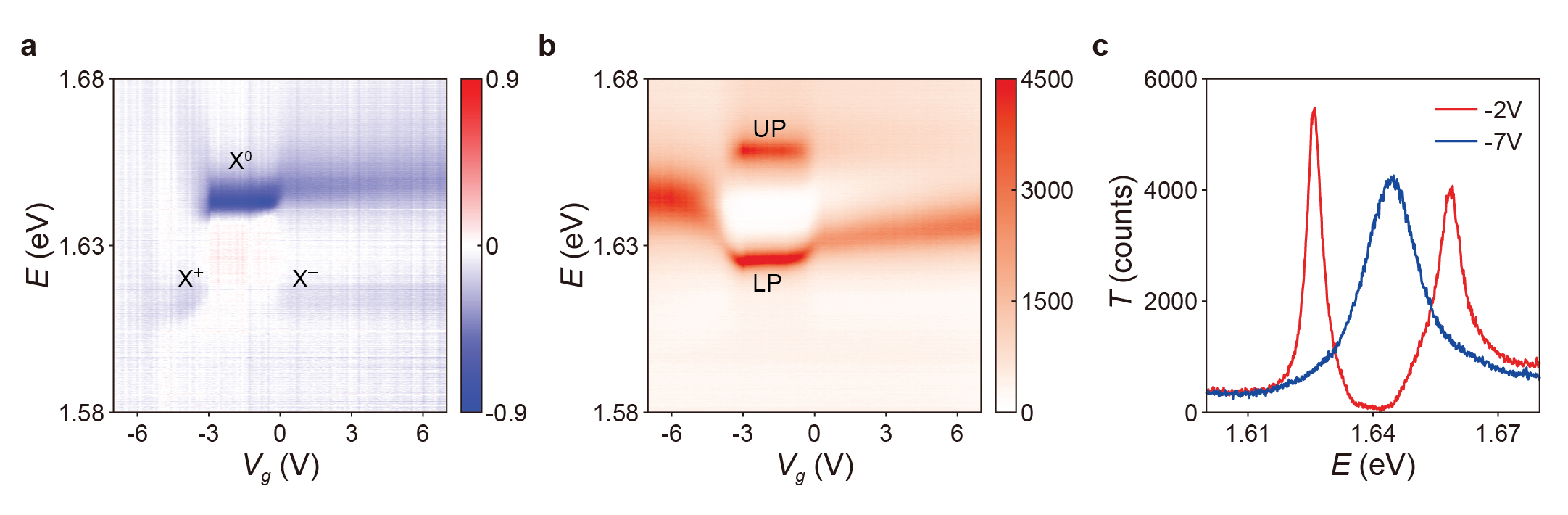}  
    \caption{\label{fig:wide} Electrical control of excitons and polaritons through electrostatic gating. a, Gate-dependent reflectance spectrum exhibits exciton resonance in the charge neutral regime from \textcolor{black}{-3V $\sim$ 0 V} and trion resonance in charge doped regime. The color plot is normalized to the reflectance spectrum at $V_g$ = -7 V, where no spectral features are observed. 
    b, Gate-dependent transmission spectrum measured through grating couplers.
    c, Transmission spectra measured in the charge neutral (red) and hole doped regimes (blue), exhibiting strong and weak couplings between excitons and cavity resonance, respectively.}
    \end{figure*}

\textcolor{black}{The hybrid device is measured at 70K, where the optical cavity is nearly on-resonance with the MoSe$_2$ excitons.}
We first assess the optical properties of MoSe$_2$ excitons and their gate dependence through reflectance measurements.
In the confocal microscope setup, the nanobeam cavity is illuminated with a broadband light source from the normal direction above, and the light directly reflected back is collected. 
Notably, the normally incident laser beam only probes exciton states that are not coupled to the cavity resonance, as the latter has negligible radiative coupling with the free space modes.
The gate-dependent reflectance spectrum shown in Fig.~3a reveals complex spectral features that vary significantly with the doping level. 
At $V_g = 0$ V, 
two peaks are observed at \textcolor{black}{1.646} eV and \textcolor{black}{1.613} eV, corresponding to the excitons ($X^0$) and negative trions ($X^-$) in the MoSe$_2$ monolayer, respectively.
The observed exciton resonance displays a broader full linewidth of 7.6 meV compared to that of flakes prepared on bare Si$_3$N$_4$ substrates. This can be ascribed to an increase in inhomogeneous broadening caused by the patterned substrate (see SI for more information).
The presence of trion resonance suggests a slight n-doping in the MoSe$_2$ monolayer even at zero gate voltage.
As $V_g$ decreases, the trion resonance gradually vanishes,
whereas the exciton feature becomes more pronounced due to reduced charge screening effect. 
In this regime, the doping level in MoSe$_2$ is nearly charge neutral and its optical response is dominated by exciton resonance. 
\textcolor{black}{Further decreasing $V_g$ to below -3 V brings the MoSe$_2$ monolayer into the p-doped regime,  where the exciton resonance vanishes again and positive trion resonance ($X^+$) starts to emerge at \textcolor{black}{1.616} eV.}
This gate-dependent measurement confirms the wide tunability of exciton oscillator strength through electrostatic doping, which is consistent with previous studies on optical properties of MoSe$_2$ monolayers\cite{ross2013electrical,li2021refractive}. 
Furthermore, we note that the initial doping levels in TMD monolayers vary among devices, which necessitates the use of electrostatic gating when interfacing TMD materials with optical devices.

Next, we study how the excitons hybridize with the nanobeam resonance to form exciton-polaritons and the electrostatic control over this hybrid TMD-PhC nanocavity.
To this end, light is guided into waveguides and the transmitted light is coupled off chip, both through grating couplers. 
\textcolor{black}{In contrast to previous far-field reflectance measurement, the waveguide is coupled to the cavity resonance in the near-field, probing the polariton features via their photonic components.}
\textcolor{black}{The measured gate-dependent transmission spectrum through grating couplers is shown in Fig.~3b.  
In the charge neutral regime ($-3{\rm V}<V_{\rm g}<0{\rm V}$), where the exciton oscillator strength is maximal, the transmission spectrum features two peaks at 1.66 eV and 1.625 eV, corresponding to the upper (UP) and lower polariton states (LP), respectively. }
The large Rabi splitting of 35 meV between the two polariton states is comparable to experimental results of similar TMD exciton polariton systems. 
The measured linewidth is 4.7 meV for LP and 7 meV for UP, respectively.
On the other hand, when the system is gradually tuned into the p-doped regime ($V_g < $ -3V), the two polariton peaks merge into a single peak at $\sim$1.644 eV. 
This peak corresponds to the optical cavity resonance and indicates a weakened exciton-photon interaction strength, signifying a shift from the strong to the weak coupling regime.
This observation is consistent with the previous reflectance measurement, which also shows a reduced exciton oscillator strength with increased hole doping level.
The measured resonance linewidth of 18 meV in the p-doped regime (Fig.~3c) is broader than that of the bare cavity (Fig.~2b), possibly due to the excess optical absorption from the free carriers in the MoSe$_2$ monolayer.
A similar transition, from strong to weak coupling, is also observed with the introduction of electron doping to the MoSe$_2$ monolayer ($V_g >$ 0V).

\begin{figure}
    \centering
    \includegraphics[width=1\columnwidth]{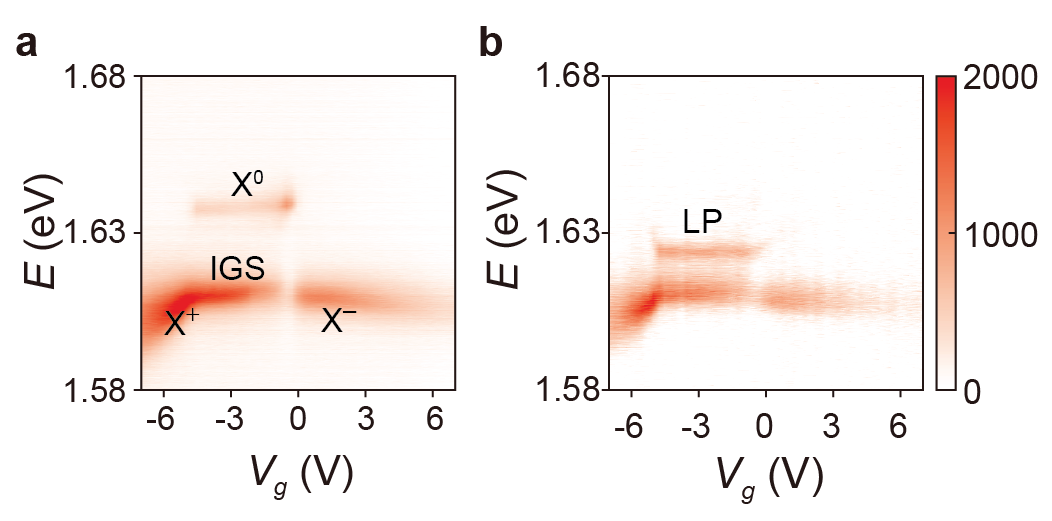} 
    \caption{\label{fig:epsart} Gate-dependent photoluminescence from the TMD-PhC nanocavity. 
    a, The PL emission collected in the normal direction as a function of gate voltage. For $V_g$ between -4.5V to 0V, we observe both neutral and in-gap exciton states (IGS). In the doped regime, the spectrum is dominated by trion emission.
    b, The PL spectrum collected through grating couplers features lower exciton polaritons in the charge neutral regime.}
\end{figure}

Finally, we characterize light emission from the TMD-PhC nanocavity by performing gate-dependent photoluminescence (PL) measurements.
We illuminate the nanocavity with a continuous-wave laser at 635nm from the normal direction and collect the PL emission in two ways: through the free-space setup in the normal direction and through grating couplers.
The PL spectrum collected from the normal direction is shown in Fig.~4a, where emission from both bare excitons and trions are observed, aligning with the previous reflectance measurement in  Fig.~3. 
\textcolor{black}{Even in the charge neutral regime ($-4.5{\rm  V} < V_{\rm g} <0 {\rm V}$), strong PL emission is observed at 1.61 eV, which is $\sim$30 meV below the exciton energy. This may be attributed to the in-gap states (IGS) resulting from the excitons bound to charged defects in MoSe$_2$\cite{kim2021free}.}
\textcolor{black}{In contrast, very different features are observed when the PL is collected using grating couplers (Fig.~4b): instead of bare excitons or trions, PL from the polariton state at 1.625 eV is observed in the charge neutral regime.}
\textcolor{black}{This PL result is also different from the previous transmission measurement (Fig.~3b), as only the lower polariton state is observable in PL --- but not the upper polariton state --- due to its energy being the lowest.}
We note that the charge neutral regime observed in the PL measurement spans from \textcolor{black}{-4.5 V to 0 V}, which is slightly broader than that from the reflectance and transmission measurements.
This discrepancy is attributed to the excess carriers induced by the pump laser, which can be mitigated by lowering the pump power. 

Previous research has studied the modulation of optical properties of TMDs through electrostatic gating\cite{ross2013electrical,li2021refractive} and its practical applications for optical modulators\cite{datta2020low,li2023purcell}.
\textcolor{black}{Our work takes one step further by venturing into the domain of strong light-matter interaction, showcasing the ability to manipulate coherent exciton-photon coupling within a nanocavity.}
This extension highlights TMDs as a versatile platform for the investigation of exciton polariton physics, further unleashing the potential for employing 2D TMDs for optoelectronic applications. 
\textcolor{black}{The compact nature of the polariton modes we have demonstrated is of particularly interest for achieving substantial polariton-polariton interactions, which are essential for nonlinear polaritonic applications. One such example is the polariton blockade, where the system's response is acutely sensitive to the number of polaritons due to strong nonlinearity.} 
The future integration of  synthetic quantum materials with intrinsically strong exciton interactions, such as twisted moiré heterostructures, with nanophotonic structures could further boost the polariton nonlinearity to the single particle level, opening the path to emergent correlated quantum phenomena.  
Moreover, our device architecture can be readily deployed for the investigation of exciton-polariton lasing by further improving the spectral properties of MoSe$_2$\cite{zhou2020controlling} and utilizing high-Q cavities.
Such exciton polariton lasing can potentially lower the lasing threshold by orders of magnitude compared to its photonic counterparts\cite{wu2015monolayer,ye2015monolayer}.

\begin{acknowledgments}
We thank Adina Ripin and Mo Li for their assistance in preparing the 2D stack. This work was partly supported by the US Office of Naval Research (ONR) through grant N00014-20-1-2325 on Robust Photonic Materials with High-Order Topological Protection and grant N00014-21-1-2703, as well as the Sloan Foundation.
\end{acknowledgments}

\bibliography{apssamp}

\end{document}